\documentclass[12pt]{article}
\usepackage{epsfig,float,cite}
\usepackage{multicol}
\usepackage{graphicx}
\usepackage{here}

\newcommand{\bea}{\begin{eqnarray}}
\newcommand{\eea}{\end{eqnarray}}

\def \beq{\begin{equation}}
\def \eeq{\end{equation}}

\def \bit{\begin{itemize}}
\def \eit{\end{itemize}}

\def\plb#1#2#3{    {\it Phys. Lett. }{\bf B #1} (#2) #3}

\def\prd#1#2#3{    {\it Phys. Rev. }{\bf D #1} (#2) #3}

\begin{document}
\begin{titlepage}

\begin{flushright}
CERN-PH-TH/2005-079\\
SLAC-PUB-11177
\end{flushright}

\centerline{\Large\bf Reduction of Charm Quark Mass Scheme Dependence}
\vspace{0.2cm}
\centerline{\Large\bf in $\bar B \rightarrow X_s \gamma$ at the NNLL Level}

\vspace{2.5cm}

\centerline{{\large\bf H.M. Asatrian$^{a}$, C. Greub$^{b}$, 
A. Hovhannisyan$^{a}$,}}
\vspace{0.5cm}
\centerline{{\large \bf T. Hurth\footnote{Heisenberg Fellow}$^{c,d}$, 
and V. Poghosyan$^{a}$}}
 
\vspace{0.1cm}

\begin{center}

{\it ${}^a$~Yerevan Physics Institute, 375036, Yerevan,
Armenia}\\

{\it ${}^b$~Institute for Theoretical Physics, University Berne, CH-3012
Berne, Switzerland}\\

{\it ${}^c$~Theoretical Physics Division, CERN, CH-1211 Geneva 23,
Switzerland}\\

{\it ${}^d$~SLAC, Stanford University, Stanford, CA 94309, USA}\\

\vspace{0.3cm}

\end{center}

\medskip

\begin{abstract}
The uncertainty of the theoretical prediction of the
$\bar B \rightarrow X_s \gamma$ branching ratio
at NLL level is dominated by the charm mass renormalization
scheme ambiguity.
In this paper we calculate those NNLL terms which are related to
the renormalization of $m_c$, in order to get an estimate of the
corresponding uncertainty at the NNLL level.
We find that these terms significantly reduce (by typically a factor of two)
the error on $\mbox{BR}(\bar B \to X_s \gamma)$ induced by the 
definition of $m_c$.
Taking into account the experimental accuracy of around $10\%$ and 
the future prospects of the $B$ factories,
we conclude that a NNLL calculation would increase the sensitivity 
of the observable
$\bar B \rightarrow X_s \gamma$ to possible new degrees
of freedom beyond the SM significantly.
\end{abstract}

\end{titlepage}

\section{Introduction}
\label{sec:intro}
The branching ratio of $\bar B \rightarrow X_s \gamma$ is a  very
sensitive probe
for new degrees of freedom beyond the standard model (SM) (for a review, see
\cite{Hurth:2003vb}).
Within supersymmetric
extensions of the SM for example, one can derive
stringent bounds on the parameter space of these models
\cite{Bertolini:1986tg,Degrassi:2000qf,Carena:2000uj,D'Ambrosio:2002ex,Borzumati:1999qt,Besmer:2001cj,Ciuchini:2002uv}.
Clearly, such bounds will be most valuable when the general nature of the
new physics beyond the SM will be identified at the forthcoming 
LHC experiments.  

Because of  the heavy mass expansion that is valid for inclusive decay modes,
the decay rate of $\bar B \rightarrow X_s \gamma$ is dominated
by the perturbatively calculable partonic decay rate
$\Gamma (b \rightarrow X_s \gamma$). QCD corrections to the
latter, due to hard-gluon exchange, are the most important perturbative
contributions; they  were calculated in the past up to the next-to-leading
logarithmic (NLL) level
\cite{Adel,GHW,Mikolaj,GH,Burasnew,Paolonew,Buras:2002tp,Asatrian:2004et,Ali:1990tj,Pott:1995if}. 
Subsequently, also electroweak corrections were calculated
\cite{Czarnecki:1998tn,Kagan:1998ym,Baranowski:1999tq,Gambino:2000fz}.
After completion of these computations, it was generally believed
that the theoretical uncertainty of the branching ratio is below
$10\%$. 

However, as first pointed out in 2001 in \cite{Gambino:2001ew},
there is an additional uncertainty in the NLL results for
$\Gamma(b\to X_s \gamma)$ which is related to the definition 
(renormalization scheme) of the charm quark mass.
Technically, the charm quark mass depencence enters through the
matrix elements $\langle s \gamma |O_{1,2}| b \rangle$ which in the
context of a NLL have to be calculated up to $O(\alpha_s)$.
As these matrix elements vanish at the lowest order,
the charm quark $m_c$ only enters (through the ratio $m_c/m_b$) 
at $O(\alpha_s)$. As a consequence, the
charm quark mass does not get renormalized in a NLL calculation,
which means that the symbol $m_c$ can be identified with $m_{c,\rm{pole}}$ or
with the $\overline{\mbox{MS}}$ mass $\bar{m}_c(\mu_c)$ at some scale $\mu_c$
or with some other definition of $m_c$. Formally, all these assignements are equivalent,
as they lead to differences which are of order $\alpha_s^2$. 

Note that in contrast to the $c$-quark mass the $b$-quark mass does get renormalized 
in a NLL calculation and we choose to express all the following results in terms of
$m_{b,\rm{pole}}$. In this respect we do not follow ref. \cite{Gambino:2001ew}, where
the $m_{b,1S}$ mass was used. Unless stated otherwise, the symbol $m_b$ stands 
for $m_{b,\rm{pole}}$ in all the formulas in this paper. 
Numerically, we use $m_b=4.8$ GeV throughout.

Numerically, it turns out that the NLL result for $\Gamma(b \to X_s \gamma)$ strongly depends
on which mass definition of the charm quark mass is used in the NLL expressions. 
To illustrate this, we first identify $m_c$ with $m_{c,\rm{pole}}$ as it was
done in all analyses before the paper of Gambino and Misiak
\cite{Gambino:2001ew}. Numerically, we 
use $m_{c,\rm{pole}}/m_{b,\rm{pole}}=0.29$ which is based on the mass difference
$m_{b,\rm{pole}}-m_{c,\rm{pole}}=3.4$ GeV fixed through the heavy mass
expansion of $m_B$ and $m_D$ and $m_{b,\rm{pole}}=4.8$ GeV.
The corresponding branching ratio then reads \cite{Gambino:2001ew}
\begin{eqnarray}
\label{BRNLLpole}
\mbox{BR}[\bar{B} \to X_s \gamma]_{E_\gamma>m_b/20} &=& 3.35 \times 10^{-4}  \, .
\end{eqnarray}
As the charm quarks which are propagating in a loop have a typical virtuality
of $m_b/2$, the authors of Ref.~\cite{Gambino:2001ew}
suggested to use $\bar{m}_c(\mu_c)$ with
$\mu_c \in [m_c,m_b]$ instead of $m_{c,\rm{pole}}$. 
A typical value for the corresponding ratio is $\bar{m}_c(\mu_c)/m_{b,\rm{pole}}=0.22$.
Using this value, the branching
ratio gets increased w.r.t. (\ref{BRNLLpole}) by about $11\%$
\cite{Gambino:2001ew}:
\begin{eqnarray}
\label{BRNLLms}
\mbox{BR}[\bar{B} \to X_s \gamma]_{E_\gamma>m_b/20} &=& 
3.73 \times 10^{-4} \, .
\end{eqnarray}
In a recent theoretical update of the NLL prediction of this branching ratio,
the uncertainty related to the definition of $m_c$ was taken into account
by varying $m_c/m_b$ in the conservative range $0.18 \le m_c/m_b \le 0.31$
which covers both, the pole mass (with its numerical error) value and the running mass
$\bar{m}_c(\mu_c)$ value with $\mu_c \in [m_c,m_b]$ \cite{Hurth:2003dk}:
\begin{equation}
\label{hurth_lunghi}
\mbox{BR}[\bar{B} \to X_s \gamma] = (3.70 \pm 0.35|_{m_c/m_b} \pm
0.02|_{\rm{CKM}} \pm 0.25|_{\rm{param.}} \pm 0.15|_{\rm{scale}}) \times 10^{-4} \, .
\end{equation}

There exists a large number of measurements of the inclusive 
 decay $\bar B \rightarrow X_s \gamma$ 
\cite{aleph,belle,cleobsg,babar1,babar2,Koppenburg:2004fz} 
and the present experimental accuracy has reached the $10 \%$ level
\cite{Alexander:2005cx}:
\beq
\mbox{BR}[\bar B \to X_s \gamma] = (3.52 \pm 0.30) \times 10^{-4} \,.
\label{world}
\eeq
In the near future, more precise data on this mode are expected from the $B$
factories. Thus, it is mandatory to reduce the present theoretical uncertainty 
accordingly. A systematic improvement certainly consists in performing a
complete NNLL calculation . 
This is, however, a very complicated task (for discussion and some results
see \cite{Misiak:2003zp,Bieri:2003ue,Gorbahn:2004my,Gorbahn:2005sa})
and a certain motivation is needed to enter such an enterprise. In the present
paper we try to give such a motivation: By
calculating those NNLL terms which are induced by
renormalizing the charm quark mass in the NLL expressions, i.e. those
terms which are sensitive to the definition of the charm quark mass,
we show that the large error at the NLL level
related to the $m_c$ definition gets significantly reduced. 
As this error is
the dominant one at the NLL level (see eq. (\ref{hurth_lunghi})), we conclude
that a complete NNLL calculation will drastically improve the theoretical 
prediction of the branching ratio.
We stress here that in the present paper we only make a statement about 
the reduction of the error at the NNLL level, and not about the central value 
of the branching ratio; this remains the topic of a complete NNLL calculation! 
 
The remainder of this paper is organized as follows. In section 
\ref{sec:method} 
we discuss in some detail how to calculate the NNLL terms induced by 
renormalizing $m_c$ in the NLL results. In order to make the paper
self-contained, we first list in section \ref{sec:analresults} 
the structure of the  NNL results
and then we present the analytical results for the
new terms discussed in section \ref{sec:method}. 
Finally, in section \ref{sec:numerics}, we numerically
investigate by how much the error related to the definition of $m_c$ gets 
reduced at the NNLL level.
\section{NNLL terms related to $m_c$ renormalization}
\label{sec:method}
As already explained in the introduction, the matrix elements
$M_{1,2}^{\rm{virt}}(m_c) = \langle s \gamma|O_{1,2}(\mu_b)|b \rangle$ only start at order 
$O(\alpha_s^1)$, or, in other words at the NLL 
order\footnote{In the present paper we use the operator basis as first introduced
in ref. \cite{Mikolaj}. $\mu_b$ denotes the renormalization scale of
$O(m_b)$.}. As a consequence, 
the definition of $m_c$ is
not fixed at this order, because $m_c$ does not get renormalized. 
This is also true for the bremsstrahlung contributions 
$M_{1,2}^{\rm{brems}}(m_c)=\langle s \gamma g|O_{1,2}(\mu_b)|b \rangle$, 
which are needed up to $O(g_s)$
for a NLL calculation.
In this section we concentrate on the virtual terms 
$M_{1,2}^{\rm{virt}}(m_c)$, as the
extension to the bremsstrahlung contributions $M_{1,2}^{\rm{brems}}(m_c)$ is
straightforward.

\begin{figure}[H]
\begin{center}
    \includegraphics[width=6.0cm]{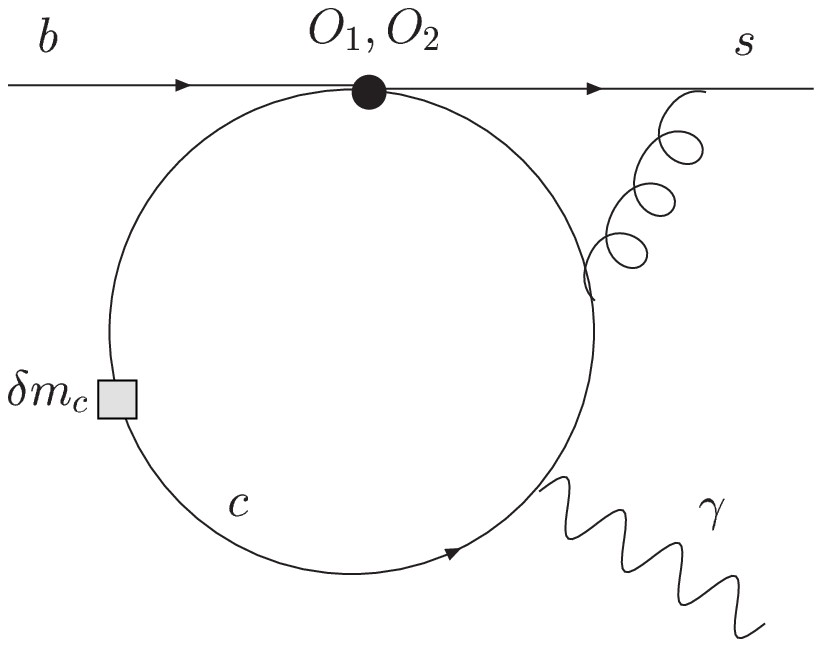}
    \includegraphics[width=6.0cm]{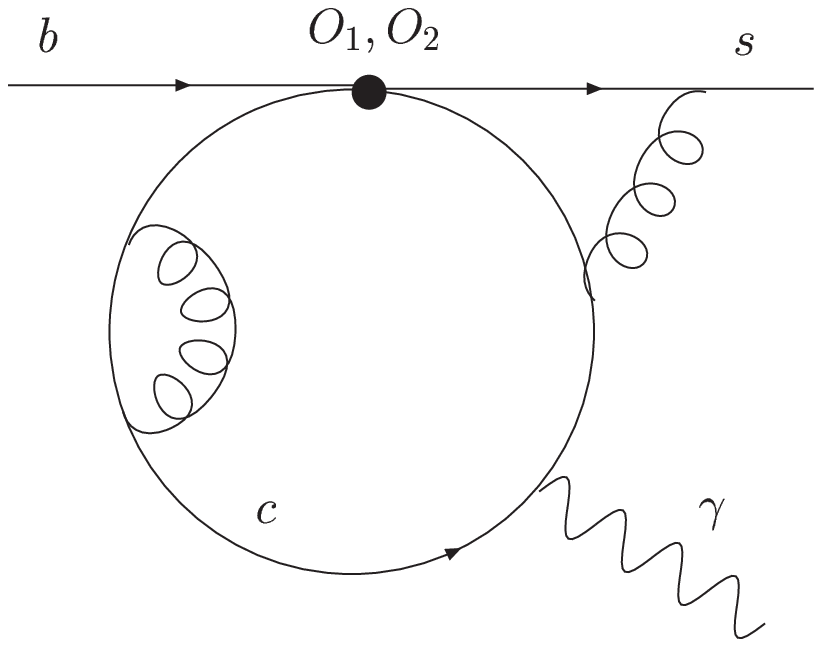}
    \vspace{0.5cm}
    \caption[]{Left frame: 
    Typical $\delta m_c$ insertion diagram (see text).    
    Right frame: 
    Typical diagram with a charm quark self-energy insertion (see text).} 
    \label{fig:diag}
    \end{center}
\end{figure}
When going to NNLL precision, the matrix elements $M_{1,2}^{\rm{virt}}(m_c)$ are needed to
$O(\alpha_s^2)$. At this level, there are -- among many other diagrams --
counterterm contributions to these
matrix elements, induced by the renormalization of $m_c$ 
(see the left frame of Fig. \ref{fig:diag}). 
The complete set of
such diagrams is generated by inserting the operator $-i\delta m_c
\bar{\psi}_c \, \psi_c$ in the $O(\alpha_s)$ diagrams of
$O_{1,2}$ in all possible ways. The sum 
$\delta M^{\rm{virt} (\epsilon)}_{1,2}(m_c) 
\cdot \delta m_c$ of all these
insertions can be obtained by replacing $m_c \to m_c + \delta m_c$ in the
$O(\alpha_s^1)$ results $M^{\rm{virt}(\epsilon)}_{1,2}(m_c)$, followed by expanding in $\delta
m_c$ up to linear order:
\begin{equation}
\label{massshift}
M_{1,2}^{\rm{virt}(\epsilon)}(m_c+\delta m_c) = M_{1,2}^{\rm{virt}(\epsilon)}(m_c)+
\delta M^{\rm{virt}(\epsilon)}_{1,2}(m_c) \cdot \delta m_c +
O((\delta m_c)^2) \, .
\end{equation} 
As $\delta m_c$ is ultraviolet
divergent, the matrix elements $M_{1,2}^{\rm{virt}(\epsilon)}(m_c)$ 
are needed in our application up to order $\epsilon^1$, as indicated by the notation in eq. 
(\ref{massshift}). 

The explicit shift $\delta m_c$ depends of course on the renormalization
scheme. When aiming at expressing the results for 
$M_{1,2}^{\rm{virt}(\epsilon)}(m_c)$ in terms
of $\bar{m}_c(\mu_b)$, the shift reads ($C_F=4/3$)
\begin{eqnarray}
\label{mcshift_bar}
\delta \bar{m}_{c}(\mu_b) &=& -\frac{\alpha_s(\mu_b)}{4\pi} \, C_F \,
\frac{3}{\epsilon} \, \bar{m}_{c}(\mu_b) \, . \nonumber
\end{eqnarray}
On the other hand,
when the result is expressed in terms of $m_{c,\rm{pole}}$, the shift reads
\begin{eqnarray}
\label{mcshift_pole}
\delta m_{c,\rm{pole}} &=& -\frac{\alpha_s(\mu_b)}{4\pi} \, C_F \, \left( \frac{3}{\epsilon}
  + 3 \ln \frac{\mu_b^2}{m_c^2} + 4 \right) \, m_{c,\rm{pole}} \, . \nonumber
\end{eqnarray}
The infinities induced by the $1/\epsilon$ terms in $\delta m_c$ get cancelled
in a full NNLL calculation, in particular by self-energy diagrams depicted in the right
frame of Fig. \ref{fig:diag}. As we do not perform a full NNLL calculation,
we suggest to consider self-energy insertions, where the self-energy $\Sigma(p^2)$
is replaced by $\Sigma_1(p^2=m_c^2)$. 
 
The $\Sigma_1$-part of the self-energy $\Sigma_1(p^2)$ 
is defined through the decomposition of the full unrenormalized self-energy 
$\Sigma(p^2)$ as    
\begin{eqnarray}
\Sigma(p^2) \equiv \Sigma_2(p^2) (\slash \hspace{-2.5mm} p - m_c) + \Sigma_1(p^2) . 
\label{splitting}
\end{eqnarray}
At the one-loop level, the corresponding pieces $\Sigma_1^R$ and
$\Sigma_2^R$ of the renormalized self-energy are  
\begin{equation}
\Sigma_2^{R}(p^2) = \Sigma_2(p^2) + \delta Z_c,\quad \quad 
\Sigma_1^R(p^2)  = \Sigma_1(p^2) + \delta m_c \, ,
\label{gaugeinvariantsplit}
\end{equation}
where $Z_c=1+\delta Z_c$ denotes the wave function renormalization constant 
of the charm quark.
Eq. (\ref{gaugeinvariantsplit}) implies that the sigularities in $\delta
M_{1,2}^{\rm{virt}(\epsilon)}(m_c) \cdot \delta m_c$ cancel when combined with the diagrams
with $\Sigma_1(p^2)$ insertions.
However, for general $p^2$, the function $\Sigma_1(p^2)$ depends on the gauge 
parameter $\xi$:
\begin{eqnarray}
\Sigma_1(p^2) = \frac{\alpha_s(\mu_b)}{4\pi} C_F m_c \left\{ 3 \left[
\frac{1}{\epsilon} + \ln \frac{\mu_b^2}{m_c^2} \right] +4
+ \left( 1-\frac{m_c^2}{p^2}\right) \left[\xi - \left( 3-\xi\frac{m_c^2}{p^2}
\right) \ln \left( 1- \frac{p^2}{m_c^2} \right)\right]
\right\} \, .
\nonumber
\end{eqnarray}
As for $p^2=m_c^2$ the self-energy piece $\Sigma_1(p^2=m_c^2)$ is
gauge-independent, we add $\Sigma_1(p^2=m_c^2)$ insertions
to $\delta M_{1,2}^{\rm{virt}(\epsilon)}(m_c) \cdot \delta m_c$.

These momentum independent
$\Sigma_1(p^2=m_c^2)$ insertions can be straightforwardly aborbed into $\delta m_c^{\rm{eff}}$
insertions:
\begin{eqnarray}
\label{mcshifts}
\delta m_{c,\rm{pole}}^{\rm{eff}} &=& \Sigma_1(p^2=m_c^2) + \delta
m_{c,\rm{pole}} = 0 \, , \\
\delta \bar{m}_{c}^{\rm{eff}}(\mu_b) &=& \Sigma_1(p^2=m_c^2) +\delta \bar{m}_c(\mu_b) =
\frac{\alpha_s(\mu_b)}{4\pi} \, C_F \, \left(3 \ln \frac{\mu_b^2}{m_c^2} + 4
\right) \, \bar{m}_c(\mu_b) \, . \nonumber
\end{eqnarray}
Finally, if we wish to express the matrix elements $M^{\rm{virt}(\epsilon)}_{1,2}(m_c)$ in terms
of $\bar{m}_c(\mu_c)$, the shift reads
\begin{equation}
\label{mcshiftsnew}
\delta \bar{m}_{c}^{\rm{eff}}(\mu_c) =
\frac{\alpha_s(\mu_b)}{4\pi} \, C_F \, \left(3 \ln \frac{\mu_c^2}{m_c^2} + 4
\right) \, \bar{m}_c(\mu_c) \, .
\end{equation}
\section{Analytical results}
\label{sec:analresults}
Before turning to the contributions induced through the renormalization
of the charm quark mass, which are NNLL terms, we first summarize
the structure of the NLL result for the branching ratio for 
$b \to X_s \gamma$.
We write the decay width for $b \to X_s\gamma$ using
a photon energy cut $E_{0}=\frac{m_b}{2}(1-\delta)=E_{max}(1-\delta)$
as
\begin{eqnarray}
\label{eq:GXsgamma}
\Gamma(b\to X_s\gamma)_{E_{\gamma}\geq E_{0}}=\Gamma(b\to s\gamma)+
\Gamma(b\to s\gamma g)_{E_{\gamma}\geq E_{0}}\,,
\end{eqnarray}
where the two parts are defined as follows:
\begin{eqnarray}
\label{eq:Gsgamma}
\Gamma(b\to s\gamma)=\frac{G_F^2}{32\pi^4}|V_{ts}^*V_{tb}|^2
\alpha_{em}m_{b,\rm{pole}}^5 \, |D|^2,
\nonumber
\end{eqnarray}
\begin{eqnarray}
\label{eq:Gsgammag}
\Gamma(b\to s\gamma g)_{E_{\gamma}\geq E_{0}}=
\frac{G_F^2}{32\pi^4}|V_{ts}^*V_{tb}|^2
\alpha_{em}m_{b,\rm{pole}}^5 \, A,
\nonumber
\end{eqnarray}
\begin{eqnarray}
\label{eq:Dvirt}
D=C_7^{(0)\rm{eff}}(\mu_b)+\frac{\alpha_s(\mu_b)}{4\pi}\left(C_7^{(1)\rm{eff}}(\mu_b)+
\sum_{i=1}^8C_i^{(0)\rm{eff}}(\mu_b)\left[r_i+
\gamma_{i7}^{(0)\rm{eff}}
\ln\left(\frac{m_b}{\mu_b}\right)\right] 
- \frac{16}{3} C_7^{(0)\rm{eff}}(\mu_b) \right), \nonumber \\
\end{eqnarray}
\begin{eqnarray}
\label{eq:Abrem}
A=\left(e^{-\alpha_s(\mu_b) \ln(\delta)[7+2\ln(\delta)]/(3\pi)}-1\right)
|C_7^{(0)\rm{eff}}(\mu_b)|^2+
\frac{\alpha_s(\mu_b)}{\pi}\sum_{i,j=1,i\leq
j}^{8} C_i^{(0) \rm{eff}}(\mu_b) C_j^{(0) \rm{eff}}(\mu_b) f_{ij}(\delta) \,.
\nonumber \\
\end{eqnarray}
The expressions for the Wilson coefficients $C_i(\mu_b)$ can be found
in \cite{Bobeth}. Their numerical values we take from table 5.1
in ref. \cite{ABGW}.
Writing the results in this specific form,
the functions
$f_{ij}(\delta)$ and $r_i$ are understood to be taken from
\cite{Mikolaj} and not from the original paper \cite{GHW}
where the results were parametrized differently.

Following common practice, we write 
the branching ratio (without taking into account non-perturbative 
corrections) as
\begin{equation}
\label{brexpr}
BR(b \to X_s\gamma)_{E_{\gamma}\geq E_{0}}=
\frac{\Gamma(b\to X_s\gamma)_{E_{\gamma}\geq E_{0}}}
{\Gamma(b \to X_c e\bar{\nu})} \, \mbox{BR}_{sl}^{exp} \, ,
\end{equation}
where the semileptonic decay rate is given by 
\begin{equation}
\label{widthsl}
\Gamma(b \to X_c e^- \bar{\nu}) = \frac{G_F^2 \, m_{b,\rm{pole}}^5}{192 \pi^3}
\, |V_{cb}|^2 \, g \! \left( \frac{m_{c}^2}{m_{b}^2} \right) \,
K \! \left( \frac{m_c^2}{m_b^2} \right) \, .
\end{equation}
$g(z)=1-8 \,z +8 \, z^3 - z^4 -12 \, z^2 \, \ln(z) \, $
is the phase-space factor and the function
\begin{eqnarray}
K(z) = 1 - \frac{2 \alpha_s(m_b)}{3\pi} \, \frac{f(z)}{g(z)},
\nonumber
\end{eqnarray}
with
\begin{eqnarray}
\label{ffun}
&&f(z) = -(1-z^2) \, \left( \frac{25}{4} - \frac{239}{3} \, z +
\frac{25}{4} \, z^2 \right) + z \, \ln(z) \left( 20 + 90 \, z
-\frac{4}{3} \, z^2 + \frac{17}{3} \, z^3 \right)
\nonumber \\
&& + z^2 \, \ln^2(z) \, (36+z^2)
+ (1-z^2) \, \left(\frac{17}
{3} -\frac{64}{3} \, z + \frac{17}{3} \, z^2
\right) \, \ln (1-z) \nonumber \\
&& -4 \, (1+30 \, z^2 + z^4) \, \ln(z) \ln(1-z)
-(1+16 \, z^2 +z^4)  \left( 6 \, \mbox{Li}(z) - \pi^2 \right)
\nonumber \\
&& -32 \, z^{3/2} (1+z) \left[\pi^2 - 4 \, \mbox{Li}(\sqrt{z})+
4 \, \mbox{Li}(-\sqrt{z}) - 2 \ln(z) \, \ln \left(
\frac{1-\sqrt{z}}{1+\sqrt{z}} \right) \right] \, .
\nonumber
\end{eqnarray}
accounts for $O(\alpha_s)$ QCD corrections. We note that $m_c$ is understood
to be the pole mass in eq. (\ref{widthsl}).

We now turn to that part of NNLL corrections which
is responsible for the reduction of the charm quark mass renormalization
scheme dependence, as explained in section \ref{sec:method}. We first turn
to terms $\delta M_{1,2}^{\rm{virt}(\epsilon)}$
induced by $m_c$ renormalization in the matrix elements
$M_{1,2}^{\rm{virt}(\epsilon)}$. To this end, we need $M_{1,2}^{\rm{virt}(\epsilon)}$
up to oder $\epsilon^1$.
In \cite{GHW} have calculated these matrix elements up
to terms $\epsilon^0$, using Mellin-Barnes
representations for generalized propogator to obtain analytic results 
in the form of the series in $z=\left(m_c/m_b\right)^2$ and 
$L=\ln(z)$. As in these calculations the expansion in $\epsilon$
was the last step, it is straightforward to calculate
$M_{1,2}^{\rm{virt}(\epsilon)}$ up to order $\epsilon^1$.

In order to get finite results for these matrix elements, we add
counterterms related to operator mixing as in ref. \cite{GHW}, adapted
however, to the operator basis defined in ref. \cite{Mikolaj}. This step
leads to $M_{1,2}^{\rm{virt,ren}}$, which we decompose as in ref. \cite{GHW}:
\begin{equation}
\label{eq:r12def}
M_{2}^{\rm{virt,ren}}= \langle s \gamma|O_7|b \rangle \, 
\frac{\alpha_s}{4 \pi} \left(
\frac{416}{81}\ln \frac{m_b}{\mu_b} 
-\frac{784}{81} \epsilon \ln^2 \frac{m_b}{\mu_b}
-4 \epsilon \ln \frac{m_b}{\mu_b} r_2^{(0)} 
+ r_2^{(0)}
+ r_{2}^{(1)} \epsilon \right) \, .
\end{equation}
We obtain for $r_2=r_2^{(0)}+\epsilon r_2^{(1)}$ (note that
$r_1=-\frac{1}{6} \, r_2$):
\begin{eqnarray}
\label{eq:virt22} 
r_2^{(0)} \hspace{-6pt} &=& \hspace{-6pt} -{\frac
{1666}{243}}-{\frac {8}{27}}\left (-48-3\,{L}^{2}-{L}^{ 3}+5\,{\pi
}^{2}+9\,L\, (-4+{\pi }^{2})+36\,\zeta (3)\right)z
+{\frac {32}{27}}\,{\pi }^{2}{z}^{3/2} \nonumber \\
&& \hspace{-6pt} + {\frac {8}{ 27}}\,\left
({L}^{3}-6\,L (-2+{\pi }^{2})+18+2\,{\pi }^{2} \right
){z}^{2}-{\frac {4}{81}}\,\left (9-182\,L+126\,{L}^{2}+14\,{ \pi
}^{2}\right ){z}^{3} \nonumber \\
&& \hspace{-6pt} -\frac{8\,i\,\pi}{27} \left[ {\frac {10}{3}} + 2\,\left
(-15-3\,L-3\,
{L} ^{2}+{\pi }^{2}\right )z+2\,\left (-3\,{L}^{2}+{ \pi }^{2}\right )
{z}^{2}
+{\frac {8}{3}}\,\left (-7+3\,L\right )\,{z}^{3} \right] \nonumber \\
\end{eqnarray}
\begin{eqnarray}
\nonumber r_2^{(1)} \hspace{-6pt} &=& \hspace{-6pt} -{\frac
{19577}{729}}+{\frac {184}{243}}\,{\pi }^{2} 
-{\frac{2}{405}} \left( -18180+75 \,{L}^{4}+3240\,{\pi }^{2}+46\,{\pi
}^{4}
- 30\,{L}^{2} (-24+7\,{ \pi }^{2}) \right. \nonumber \\ 
&& \hspace{-6pt}\left. + 9000\,\zeta(3)+120\,L 
(-66+14\,{\pi }^{2}+27\, \zeta (3) ) 
\right) z  
 - {\frac {32}{81}}\,{\pi }^{2}\left (-49 +6\,L+24\,\ln
(2)\right ){z}^{3/2} \nonumber \\
&& \hspace{-6pt} + {\frac {2}{81}}\left( 48\,{L}^{3}-
15\,{L}^{4}+24\,{L}^{2} (-3+{\pi }^{2}) - 24\,L (3+5\,{ \pi }^{2})+
1116+36\,{\pi
}^{2} \right. \nonumber \\ 
&& \hspace{-6pt} \left. + 40\,{\pi
}^{4}+432\,\zeta (3) \right ){z}^{2}-{\frac {1120}{81} }\,{\pi
}^{2}{z}^{5/2} 
 +{\frac {1}{729}}\left( 22705-2484\,
{L}^{2}+4536\,{L}^{3} -6036\,{\pi }^{2} \right. \nonumber \\ 
&& \hspace{-6pt}
\left. + 6\,L  (-1783+192\,{\pi
}^{2 })
+8208\,\zeta (3)\right ){z}^{3} \nonumber  \\
 && \hspace{-6pt} +{\frac {8\,i\,\pi}{27}} \,\left[
-{\frac {221}{9}}+\left (15\,{L}^{2}-6\,{L}^{3}-4\,L (-9+{\pi
}^{2} )+186-10 \,{\pi }^{2}-36\,\zeta (3)\right) z \right. \nonumber \\
&& \hspace{-6pt} 
- 2\left( -3
-6\,{L}^{2}+3\,{L}^{3}+
2\,{\pi }^{2}+2\,L{\pi }^{2}+
 18\,\zeta (3)\right) {z}^{2} \nonumber \\
&& \hspace{-6pt} \left. +\frac{4}{9}\, 
(-67+66\,L+ 9\,{L}^{2}+12\,{\pi }^{2} ){z}^{3} \right]
\end{eqnarray}
In these formulas we retained all terms up to order $z^3$, as higher order terms
contribute much less than $1\%$. Nevertheless,
in the numerical evaluations in section \ref{sec:numerics}
all terms up to $z^6$ were included.

At the level of the decay width, the implementation of
the contribution coming from renormalization of
the $c$-quark mass in the virtual contributions is (according to eq. 
(\ref{massshift})) most easily achieved
by replacing $r_{1,2}$ in eq. (\ref{eq:Dvirt})  
by $r_{1,2}^{(0)} + \Delta r_{1,2}$,  where
\begin{eqnarray}
\label{eq:deltar2}
\Delta r_{1,2}=\delta m_c \frac{d}{dm_c}
\left(r_{1,2}^{(0)}+\epsilon\, r_{1,2}^{(1)}\right) \, .
\end{eqnarray}

At the NLL order, the bremsstrahlung corrections to the decay
width are encoded in the quantities $f_{ij}(\delta)$ 
(see eq. (\ref{eq:Abrem})), which correspond to the interference terms
$(O_i,O_j)$.
 In the following, we calculate the shifts
$\Delta f_{ij}$ to these quantities induced by the renormalization of the
charm quark mass. In principle, we calculate the decay width using a photon
energy cut $\delta=0.9$ (see eq. (\ref{eq:GXsgamma})). 
However, as all bremsstrahlung contributions which
contain charm quark loops are finite for $\delta \to 1$, we can 
approximate these terms by putting $\delta=1$.
Numerically the relative error is of order $10^{-4}$.

We first calculate the shift $\Delta f_{27}$.
To this end, we shift the charm quark mass in the matrix element of 
$\langle s \gamma g|O_2|b\rangle$ 
as in eq. (\ref{massshift}) and then work out the interference with
$\langle s \gamma g|O_7|b\rangle$. Because of the $1/\epsilon$ term in
$\delta m_c$, the result is ultraviolet singular. In a full
NNLL calculation this singularity gets cancelled when combined
with self-energy insertions in the charm quark lines in the matrix element
of $O_2$. We therefore do the phase space integrations
involved in the derivation of $f_{27}$ (or $\Delta f_{27}$) in d=4 dimensions. 
As only the matrix element of $O_2$ depends on $m_c$, the shift $\Delta f_{27}$
can be constructed by first considering the quantity $f_{27}$ itself. 
Using the integral representation for the building
block for photon and gluon emission from the $c$-quark loop \cite{GHW},
one obtains after integration over all but one of the phase space 
parameters
\begin{eqnarray}
\label{intrep}
f_{27}=-\frac{8}{9} \left(\frac{\mu_b}{m_b} 
\right)^{2\epsilon}\int dx \, dy \, du \, u^2 (1-u) (1-x) y \,
\frac{[x (1-x)]^{-\epsilon} \, \Gamma(2+\epsilon) \,
\mbox{e}^{\epsilon \gamma+i \epsilon \pi}}{\left[u y-\frac{z}{x (1-x)}
+i \eta  \right]^{(1+\epsilon)}} \, .
\end{eqnarray}
Here $x,y$ are Feynman parameters and $u$ is the remaining phase space
parameter, $0\leq x,y,u\leq 1$. To solve the integrals,
we use the Mellin--Barnes representation for the generalized propagator
\begin{eqnarray*}
\left[u y-\frac{z}{x (1-x)}+i \eta \right]^{-1-\epsilon}
\hspace{-0.8cm}&=&\frac{1}{2i\pi\Gamma(1+\epsilon)}\int_\gamma
ds\frac{\mbox{e}^{i \pi s} \, \Gamma(-s) \, \Gamma(1+\epsilon+s)}
{(uy)^{1+\epsilon}}
\left[\frac{z}{uyx(1-x)}\right]^s
\end{eqnarray*}
appearing in eq. (\ref{intrep}).
$\gamma$ denotes the integration path parallel to imaginary axes
which hits the real axes somewhere between $(-1-\epsilon)$ and $0$.
Closing the integration path in the right $s$-half plane,
one gets an expansion for $f_{27}$ in $z=(m_c/m_b)^2$ and $L=\ln(z)$.

\noindent
The shift $\Delta f_{27}$ is then obained as 
\begin{eqnarray}
\label{eq:deltaf27}
\Delta f_{27}=\delta m_c \, \frac{dz}{dm_c} \, \frac{df_{27}}{dz} = 
2 \frac{m_c}{m_b}\frac{\delta m_c}{m_b}\left(f_{27}^{d
(0)}+
\epsilon \, f_{27}^{d(1)}\right) \, .
\end{eqnarray}
To summarize, the NNLL contributions due to renormalization of $m_c$
in the $(O_2,O_7)$ interference are taken into account
by replacing $f_{27} \to f_{27}^{(0)} + \Delta f_{27}$ 
in eq. (\ref{eq:Abrem}). Explicitly, we find: 
\begin{eqnarray}
\label{eq:brem270} \nonumber
f_{27}^{(0)}&=&-\frac{8}{9}\left[\frac{1}{12} +\frac{1}{8}(7 - 2\pi^2 
+ 6L + 2L^2) z + 
(\pi^2 - 2L - L^2) z^2 \right.\\&+&\left.\frac{1}{4}(-11 - 6\pi^2 - 
4L + 6L^2) z^3 
+ \frac{1}{3}(-6 + 10L) z^4 \right.\\
&+&\left. \frac{1}{24} (13 + 70L) z^5 + \frac{1}{15} (32 + 63 L) z^6
\nonumber 
\right] \, ,
\end{eqnarray}
\begin{eqnarray}
\label{eq:bremd270} \nonumber f_{27}^{d(0)}&=&-\frac{8}{9}\left[ \frac
{1}{8} (13-2\pi^2+10L+2L^2)+ 
     2(-1+\pi^2 - 3 L-L^2) z\right.\\ &+& \left.\frac{1}{4}(-37 - 
18\pi^2 + 18L^2) z^2 + 
     \frac{2}{3}(-7 +20L)z^3\right.\\&+&\left. 
     \frac{5}{24} (27+ 70L) z^4
\nonumber +
     \frac{1}{5} (85+126L) z^5
\right] \, ,
\end{eqnarray}

\begin{eqnarray}
\label{eq:bremd271} \nonumber f_{27}^{d(1)}&=&-\frac{8}{9}\left[ \frac
{1}{48} (165-52\pi^2-4(-15+\pi^2)L-18L^2-8L^3-72\zeta(3))\right.\\&+& \left.
\nonumber 
     \frac{1}{3}(2(-12+\pi^2)L -3 L^2+4L^3+6(-5+4\pi^2+6\zeta(3))) z
\right.\\ &-& 
     \left.\frac{3}{4}(-7+15\pi^2+(15+2\pi^2)L-15L^2+4L^3+36\zeta(3)) z^2
     \right.\\&-&\left.\nonumber 
     \frac{5}{27}(235+48\pi^2-204L+36L^2) z^3\right.\\&+&\left.
     \nonumber 
     \frac{1}{432}(-10076-4200\pi^2+19425L-3150L^2) z^4
\right.\\&+&\left.\nonumber 
     \frac{1}{250}(-3554-4200\pi^2+20685L-3150L^2) z^5
\right] \, .
\end{eqnarray}
Note that $f_{27}^0$ in eq. (\ref{eq:brem270}) is an expanded
version in $z$ of the integral expression for $f_{27}$ 
in ref. \cite{Mikolaj}.
We further note that $f_{28}=-\frac{1}{3}f_{27}$, 
          $f_{17}=-\frac{1}{6}f_{27}$, 
          $f_{18}=\frac{1}{18}f_{27}$; the same relations also hold
for the respective $\Delta f_{ij}$ (see for instance, \cite{Asatryan:2002iy}). 

Finally, we turn to the shift $\Delta f_{22}$ related to the
$(O_2,O_2)$-interference. To derive this quantity, one
has to perform the shift $m_c \to m_c + \delta m_c$ only in one of the
two interfering one-loop amplitudes
$M_2^{\rm{brems}}=\langle s \gamma g|O_2|b \rangle$. To this end, one
writes integral representations for both, 
$M_2^{\rm{brems}}$ and $dM_2^{\rm{brems}}/dm_c$. $\Delta f_{22}$ is then
represented as a five dimensional integral (4 Feynman parameters and one
phase space parameter), which can be solved by double Mellin-Barnes
techniques (see for instance \cite{Asatryan:2001zw}). 
Omitting the detail of this calculation, the terms
induced by renormalizing $m_c$ in the $(O_2,O_2)$ bremsstrahlung terms
are implemented in eq. (\ref{eq:Abrem}) by replacing    
$f_{22} \to f_{22}^{(0)}+\Delta f_{22}$,  where
\begin{eqnarray}
\label{eq:deltaf22}
\Delta f_{22}=2 \frac{m_c}{m_b}\frac{\delta m_c}{m_b}\left(f_{22}^{d
(0)}+
\epsilon \, f_{22}^{d(1)}\right) \, .
\end{eqnarray}
Explicity, we get:
\begin{eqnarray}
\label{eq:brem220} \nonumber
f_{22}^{(0)}&=&0.04938272+(16.64197 + 
       1.887290L - 0.4444444L^2 - 
       0.09876543L^3) z \\ &+& \nonumber
     (57.92026 + 47.67037L + 
       1.185185L^2 + 3.134737L^3 + 
       0.05925926L^5) z^2 \\& +& 
  (-93.12628 + 
       32.36078L - 12.95977L^2 + 
       1.777778L^3 - 0.2962963L^4) z^3 \nonumber\\ &+& 
       (11.92082 - 
       11.21491L + 2.074074L^2 - 
       0.5925926L^3) z^4\\&+& \nonumber 
        (0.6482797 - 
       4.160089L + 0.1810700L^2 - 
       0.3292181L^3) z^5\\ &+& \nonumber
        (-1.125313 - 
       4.320604L - 0.2444444L^2 - 
       0.3456790L^3) z^6 \, , 
\end{eqnarray}

\begin{eqnarray}
\label{eq:bremd220} \nonumber
f_{22}^{d(0)}&=&18.52926 + 0.9984013L - 
     0.7407407L^2 - 0.09876543L^3\\&+& \nonumber
      (163.5109 + 
       97.71112L + 11.77458L^2 + 
       6.269473L^3 + 0.2962963L^4 + 
       0.1185185L^5) z \\ &+& \nonumber
        (-247.0180 + 71.16280L - 
       33.54596L^2 + 4.148148L^3 - 
       0.8888889L^4) z^2 \\& +& 
   (36.46839 - 40.71149L + 
       6.518519L^2 - 2.370370L^3) z^3 \\ &+& \nonumber
        (-0.9186906 - 20.43831L - 
       0.08230453L^2 - 1.646091L^3) z^4 \\&+& \nonumber 
        (-11.07248 - 26.41251L - 
       2.503704L^2 - 2.074074L^3) z^5 \, ,
\end{eqnarray}

\begin{eqnarray}
\label{eq:bremd221} \nonumber
f_{22}^{d(1)}&=&41.24600 - 7.794263L - 
     0.7525535L^2 + 0.3950617L^3 + 
     0.07407407L^4  \\&+& \nonumber
      (234.4505 + 
       44.95451L - 64.68047L^2 - 
       0.5200208L^3\\ &-&\nonumber 
       4.498135L^4 - 
       0.05925926L^5 - 0.08559671L^6) z\\ &+& \nonumber
        (-368.8104 + 
       245.2526L - 95.81857L^2 + 
       28.84099L^3 - 3.851852L^4 + 
       0.5728395L^5) z^2 \\& +& 
   (-3.708986 - 
       93.16023L + 34.50854L^2 - 
       7.670782L^3 + 1.283951L^4) z^3 \\& +& \nonumber
        (-45.10910 - 
       54.73258L + 13.39517L^2 - 
       4.073160L^3 + 0.8779150L^4) z^4
       \\ &+& \nonumber
        (-92.60340 - 74.33726L + 
       10.32328L^2 - 4.786008L^3 + 
       1.094650L^4) z^5 \, . 
\end{eqnarray}
We decided to give the expansion coefficients in these equations
in numerical form, because the exact results are somewhat lenghty.
We note that $f_{22}^0$ in eq. (\ref{eq:brem220}) is an expanded
version in $z$ of the integral expression for $f_{22}$ 
in ref. \cite{Mikolaj}.
We further note that $f_{11}=\frac{1}{36}f_{22}$ and 
                     $f_{12}=-\frac{1}{3}f_{22}$; 
                     the same relations also hold
                     for the respective $\Delta f_{ij}$. 

\noindent
These analytical results are 
defined parts of the complete NNLL 
contribution which can be used within a future NNLL calculation.

\section{Numerical results}
\label{sec:numerics}
In the following analysis we show that the NNLL terms, induced through the
renormalization of $m_c$, drastically reduce the error related to the
definition of the charm quark mass in $\mbox{BR}(b \to X_s \gamma)$.
To illustrate this feature as clearly as possible, we take the fixed 
values shown in Table \ref{tab:main_inputs} for the input parameters.
\begin{table}[h]
\centering
\begin{tabular}{|c|c|c|}
\hline
   $m_b= 4.8 $~GeV    
&  $m_{c,\rm{pole}}/m_b=0.29$
&  $m_Z= 91.187$~GeV  
 \\ \hline
\end{tabular}
\vskip 0.3 cm 
\begin{tabular}{|c|c|c|c|}
\hline
   $\alpha_s(m_Z)= {0.119}$ 
&  $\alpha_{\rm em}=1/137.036$ 
&  $|V^*_{ts} V_{tb}/V_{cb}|^2=0.95$
&  $BR_{\rm{sl}}=10.49\%$
 \\ \hline
\end{tabular}
\caption{Input parameters used in the 
numerical analysis.}
\label{tab:main_inputs}
\end{table}
In particular, we use the fixed ratio $m_{c,\rm{pole}}/m_{b,\rm{pole}}=0.29$.
Furthermore, we always leave the semileptonic decay
width, which enters the
branching ratio for $b \to X_s \gamma$ through eq. (\ref{brexpr}),  
expressed in terms of $m_{c,\rm{pole}}$ as given in eq. 
(\ref{widthsl}). 
In this way the $m_c$ definition dependence of 
the $\mbox{BR}(b \to X_s \gamma)$ only comes from the numerator in
eq. (\ref{brexpr}). For our studies, we neglect electroweak 
corrections and non-perturbative effects. 
As already mentioned, in the bremsstrahlung 
contribution we use $\delta=0.9$ for the lower cut in the photon energy
(see eq. (\ref{eq:GXsgamma})). 

\begin{figure}[H]
\begin{center}
    \includegraphics[width=12.0cm]{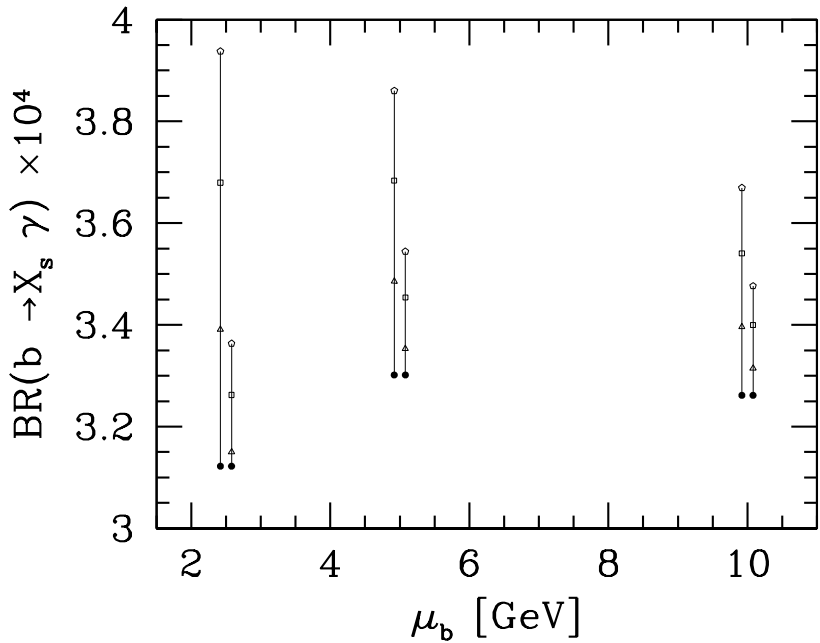}
    \caption[]{$\mbox{BR}(b \to X_s \gamma)$ for three values of $\mu_b$.
For each value of $\mu_b$ the left string shows the NLL results 
for $m_{c,pole}$ (solid dot) and for $\bar{m}_c(\mu_c)$ with
$\mu_c=1.25;2.5;5.0$ GeV (open symbols). The right strings
show the corresponding NLL results supplemented by the $\delta m_c$
mass insertions and the $\Sigma_1(p^2=m_c^2)$ insertions  
(see text for more details).} 
    \label{fig:plots}
    \end{center}
\end{figure}

Starting from $m_{c,\rm{pole}}=0.29 \cdot 4.8$ GeV = 1.392 GeV,
we first calculate $\bar{m}_c(m_{c,\rm{pole}})$, using the one-loop
expression
\begin{equation}
\label{onelooprel}
\bar{m}_c(m_{c,\rm{pole}})=m_{c,\rm{pole}} \left[ 1
-\frac{\alpha_s(m_{c,\rm{pole}})}{\pi} \, C_F \,
\right] \, .
\end{equation} 
To get $\bar{m}_c(\mu_c)$ for an arbitrary scale (typically between 1.25 GeV
and 5 GeV), we use two-loop running (with 5 flavours) according to
\begin{equation}
\bar{m}_c(\mu_c)=\bar{m}_c(\mu_0) \left(\frac{\alpha_s(\mu_c)}{\alpha_s(\mu_0)}
\right)^{\frac{\gamma_m^{(0)}}{2 \beta_0}} \left[1+\left(
\frac{\gamma_m^{(1)}}{2\beta_0}-\frac{\beta_1
  \gamma_m^{(0)}}{2\beta_0^2}\right) \, 
\frac{\alpha_s{(\mu_c)} - \alpha_s(\mu_0)}{4\pi}\right]
\end{equation}
with $\mu_0=m_{c,\rm{pole}}$. 
Numerically, we get the values shown in table \ref{tab:mc_values}.
\begin{table}[h]
\centering
\begin{tabular}{|c|c|c|}
\hline
   $\bar{m}_{c}(1.25)/m_b= 0.257 $ &    
   $\bar{m}_{c}(2.5)/m_b= 0.214 $ &    
   $\bar{m}_{c}(5.0)/m_b= 0.187 $   \\  
\hline
\end{tabular}
\caption{$\bar{m}_c(\mu_c)/m_b$ for $\mu_c=1.25,2.5,5$ GeV using
$m_{c,\rm{pole}}/m_b=0.29$ as input.}
\label{tab:mc_values}
\end{table}
In Figure \ref{fig:plots} our results are given 
for three different values of $\mu_b$, where $\mu_b$
represents the usual renormalization scale of the effective field theory. 
We compare the branching ratio for $b\to X_s\gamma$ within the pole and 
the $\overline{\rm{MS}}$ scheme for the charm quark mass.
Within each vertical string the solid dot represents the branching ratio using 
$m_{c,\rm{pole}}$, while the open symbols correspond 
to $\bar{m}_c(\mu_c)$ 
for $\mu_c=1.25$ GeV (triangle), 
    $\mu_c=2.5$ GeV (quadrangle) and
    $\mu_c=5.0$ GeV (pentagon), respectively. 
For each $\mu_b$ the left string shows the value of
the branching ratio at the NLL level, while the right string shows  
the corresponding value where in addition $\delta m_c$ mass insertions and 
$\Sigma_1(p^2=m_c^2)$ insertions were taken into account, as
explained in detail in section \ref{sec:method}.
Because the combination of these insertions is zero 
by construction for the pole scheme (see eq. (\ref{mcshifts})), 
the solid dots are at the same place in the left and the right string 
for a given value of $\mu_b$. 

From Figure \ref{fig:plots} we see that the error related to the charm 
quark mass definition gets significantly reduced when taking 
into account NNLL terms connected
with mass insertions. Taking as an example the results for $\mu_b=5$ GeV, 
we find that at the NLL level the
branching ratio evaluated for $\bar{m}_c(2.5 \ \mbox{GeV})$
is $12.6\%$ higher than the one based on
$m_{c,\rm{pole}}$, in agreement with ref. \cite{Gambino:2001ew}. 
Including the new contributions, these $12.6\%$
get reduced to $5.1\%$.   

A remark concerning the remaining NNLL terms is in order: As these terms
give contributions to the branching ratio which (up to 
terms of order $\alpha_s^3$) do not depend on charm quark mass definition,
the error related to $m_c$ in the full NNLL result is expected to stay
essentially the same as estimated in the present paper.

However, to obtain a NNLL prediction for the central value
of the branching ratio, it is of course necessary to calculate all NNLL terms. 

Summing up, we have shown that the relatively large error related to 
the definition 
of the charm quark mass in the NLL result for 
$\mbox{BR}(b \to X_s \gamma)$ gets significantly reduced (typically by a factor
of 2) at the NNLL level.
Taking into account the present experimental accuracy of around $10\%$ 
and  the future prospects of the $B$ factories and also of possible
Super-$B$ factories \cite{SuperB1,SuperB2}, 
we conclude that a future NNLL QCD calculation of the 
$b \rightarrow X_s \gamma$ branching ratio will significantly 
increase the sensitivity of this observable to possible new physics.

\section*{ Acknowledgements} 
The team from Yerevan is partially supported by the ANSEF-05-PS-hepth-813-100
and the NFSAT PH 095-02 (CRDF 12050) programs.
\newline
C.G. is partially supported by: the Swiss National Foundation; 
RTN, BBW-Contract No.~01.0357 and EC-Contract HPRN-CT-2002-00311 (EURIDICE).
\newpage

\bigskip

\frenchspacing
\footnotesize
\begin{multicols}{2}

\end{multicols}

\begin{thebibliography}{99}


\bibitem{Hurth:2003vb}
T.~Hurth,
Rev. Mod. Phys. {\bf 75} (2003) 1159
[arXiv:hep-ph/0212304].

\bibitem{Bertolini:1986tg}
S.~Bertolini, F.~Borzumati and A.~Masiero,
Phys.\ Lett.\ B {\bf 192} (1987) 437;
S.~Bertolini, F.~Borzumati, A.~Masiero and G.~Ridolfi,
Nucl.\ Phys.\ B {\bf 353} (1991) 591.

\bibitem{Degrassi:2000qf}
G.~Degrassi, P.~Gambino and G.~F.~Giudice,
JHEP {\bf 0012}, 009 (2000)
[arXiv:hep-ph/0009337].


\bibitem{Carena:2000uj}
M.~Carena, D.~Garcia, U.~Nierste and C.~E.~M.~Wagner,
Phys.\ Lett.\ B {\bf 499}, 141 (2001)
[arXiv:hep-ph/0010003].



\bibitem{D'Ambrosio:2002ex}
G.~D'Ambrosio, G.~F.~Giudice, G.~Isidori and A.~Strumia,
Nucl.\ Phys.\ B {\bf 645}, 155 (2002)
[arXiv:hep-ph/0207036].

\bibitem{Borzumati:1999qt}
F.~Borzumati, C.~Greub, T.~Hurth and D.~Wyler,
Phys.\ Rev.\ D {\bf 62}, 075005 (2000)
[arXiv:hep-ph/9911245].


\bibitem{Besmer:2001cj}
T.~Besmer, C.~Greub and T.~Hurth,
Nucl.\ Phys.\ B {\bf 609}, 359 (2001)
[arXiv:hep-ph/0105292].

\bibitem{Ciuchini:2002uv}
M.~Ciuchini, E.~Franco, A.~Masiero and L.~Silvestrini,
Phys.\ Rev.\ D {\bf 67}, 075016 (2003)
[Erratum-ibid.\ D {\bf 68}, 079901 (2003)]
[arXiv:hep-ph/0212397].

\bibitem{Adel}
K.~Adel and Y.~Yao,
\prd{49}{1994}{4945},
[arXiv:hep-ph/9308349].


\bibitem{GHW}
C.~Greub, T.~Hurth and D.~Wyler,
\plb{380}{1996}{385},
[arXiv:hep-ph/9602281];
\prd{54}{1996}{3350},
[arXiv:hep-ph/9603404].

\bibitem{Mikolaj}
K.~Chetyrkin, M.~Misiak and M.~M\"unz,
\plb{400}{1997}{206},
[arXiv:hep-ph/9612313].

\bibitem{GH}
C.~Greub and T.~Hurth,
\prd{56}{1997}{2934},
[arXiv:hep-ph/9703349].

\bibitem{Burasnew}
A.~J.~Buras, A.~Czarnecki, M.~Misiak and J.~Urban,
Nucl.\ Phys.\ B {\bf 611} (2001) 488,
[arXiv:hep-ph/0105160].

\bibitem{Paolonew}
P.~Gambino, M.~Gorbahn and U.~Haisch,
arXiv:hep-ph/0306079.

\bibitem{Buras:2002tp}
  A.~J.~Buras, A.~Czarnecki, M.~Misiak and J.~Urban,
  Nucl.\ Phys.\ B {\bf 631} (2002) 219
  [arXiv:hep-ph/0203135].

\bibitem{Asatrian:2004et}
H.~M.~Asatrian, H.~H.~Asatryan and A.~Hovhannisyan,
Phys.\ Lett.\ B {\bf 585} (2004) 263

\bibitem{Ali:1990tj}
A.~Ali and C.~Greub,
Z.\ Phys.\ C {\bf 49} (1991) 431.

\bibitem{Pott:1995if}
N.~Pott,
Phys.\ Rev.\ D {\bf 54} (1996) 938
[arXiv:hep-ph/9512252].

\bibitem{Czarnecki:1998tn}
A.~Czarnecki and W.~J.~Marciano,
Phys. Rev. Lett. {\bf 81} (1998) 277
[arXiv:hep-ph/9804252].


\bibitem{Kagan:1998ym}
A.~L.~Kagan and M.~Neubert,
Eur.\ Phys.\ J.\ C {\bf 7}, 5 (1999)
[arXiv:hep-ph/9805303].


\bibitem{Baranowski:1999tq}
K.~Baranowski and M.~Misiak,
Phys.\ Lett.\ B {\bf 483}, 410 (2000)
[arXiv:hep-ph/9907427].

\bibitem{Gambino:2000fz}
P.~Gambino and U.~Haisch,
JHEP {\bf 0009}, 001 (2000)
[arXiv:hep-ph/0007259]; \\
P.~Gambino and U.~Haisch,
JHEP {\bf 0110}, 020 (2001) 
[arXiv:hep-ph/0109058].

\bibitem{Gambino:2001ew}
P.~Gambino and M.~Misiak,
Nucl.\ Phys.\ B {\bf 611}, 338 (2001)
[arXiv:hep-ph/0104034].

\bibitem{Hurth:2003dk}
T.~Hurth, E.~Lunghi and W.~Porod,
Nucl.\ Phys.\ B {\bf 704} (2005) 56
[arXiv:hep-ph/0312260].

\bibitem{aleph}
R.~Barate {\it et al.}  [ALEPH Collaboration],
Phys.\ Lett.\ B {\bf 429} (1998) 169.

\bibitem{belle}
K.~Abe {\it et al.}  [Belle Collaboration],
Phys.\ Lett.\ B {\bf 511} (2001) 151,
[arXiv:hep-ex/0103042].

\bibitem{cleobsg}
S.~Chen {\it et al.}  [CLEO Collaboration],
Phys. Rev. Lett. {\bf 87} (2001) 251807,
[arXiv:hep-ex/0108032].

\bibitem{babar1}
B.~Aubert {\it et al.}  [BaBar Collaboration],
arXiv:hep-ex/0207074.

\bibitem{babar2}
B.~Aubert {\it et al.}  [BaBar Collaboration],
arXiv:hep-ex/0207076.


\bibitem{Koppenburg:2004fz}
  P.~Koppenburg {\it et al.}  [Belle Collaboration],
  Phys.\ Rev.\ Lett.\  {\bf 93} (2004) 061803
  [arXiv:hep-ex/0403004].


\bibitem{Alexander:2005cx}
  J.~Alexander {\it et al.}  [Heavy Flavor Averaging Group (HFAG)],
  arXiv:hep-ex/0412073.

\bibitem{Misiak:2003zp}
  M.~Misiak,
  Nucl.\ Phys.\ Proc.\ Suppl.\  {\bf 116} (2003) 279.


\bibitem{Bieri:2003ue}
  K.~Bieri, C.~Greub and M.~Steinhauser,
  Phys.\ Rev.\ D {\bf 67} (2003) 114019
  [arXiv:hep-ph/0302051].

\bibitem{Gorbahn:2004my}
 M.~Gorbahn and U.~Haisch,
 Nucl.\ Phys.\ B {\bf 713} (2005) 291
 [arXiv:hep-ph/0411071].

\bibitem{Gorbahn:2005sa}
M.~Gorbahn, U.~Haisch and M.~Misiak,
arXiv:hep-ph/0504194.

\bibitem{Bobeth}
  C.~Bobeth, M.~Misiak and J.~Urban,
  Nucl. Phys. B {\bf 574} (2000) 291
  [arXiv:hep-ph/9910220].

\bibitem{ABGW}
  H.~M.~Asatrian, K.~Bieri, C.~Greub and M.~Walker,
  Phys. Rev. D {\bf 69} (2004) 074007
  [arXiv:hep-ph/0312063].

\bibitem{Asatryan:2002iy}
H.~H.~Asatryan, H.~M.~Asatrian, C.~Greub and M.~Walker,
Phys.\ Rev.\ D {\bf 66}, 034009 (2002)
[arXiv:hep-ph/0204341].

\bibitem{Asatryan:2001zw}
H.~H.~Asatryan, H.~M.~Asatrian, C.~Greub and M.~Walker,
Phys.\ Rev.\ D {\bf 65}, 074004 (2002)
[arXiv:hep-ph/0109140].

\bibitem{SuperB1}   A.~G.~Akeroyd {\it et al.}  
[SuperKEKB Physics Working Group],
  ``Physics at super B factory,''
  arXiv:hep-ex/0406071.

\bibitem{SuperB2} 
J.Hewett {\it et al.}
``The discovery potential of a Super B Factory," SLAC  
Workshops, Stanford, USA, 2003, arXiv:hep-ph/0503261.

\end{thebibliography}
\end{document}